\documentclass[copyright]{eptcs}
\providecommand{\event}{AFL 2026}
\usepackage{latexsym}\usepackage{amssymb}

\newcommand*{\set}[1]{\{#1\}} 
\newcommand*{\st}{\,{:}\,} 
\newcommand*{\abs}[1]{|{#1}|} 
\newcommand*{\angl}[1]{\langle{}#1{}\rangle} 
\newcommand*{\ceil}[1]{\lceil{#1}\rceil} 
\newcommand*{\quo}[1]{``#1''} 
\newcommand*{\mathquo}[1]{\mbox{``}#1\mbox{''}} 
\newcommand*{\xss}[1]{_{\scriptscriptstyle{#1}}} 
\newcommand*{\xsr}[1]{_{\scriptscriptstyle\mathrm{#1}}} 
\newcommand*{\usr}[1]{^{\scriptscriptstyle\mathrm{#1}}} 
\newcommand*{\str}{^{\ast}} 
\newcommand*{\mmod}{{\,\bmod\,}} 
\newcommand*{\eps}{\varepsilon} 
\newcommand*{\emset}{\textit{\O}} 
\newcommand*{\bbn}{\ensuremath{\mathbb{N}}} 
\newcommand*{\trs}[1]{\,\raisebox{-0.3ex}[0ex]{
  $\stackrel{{#1}}{\raisebox{0ex}[0.65ex]{$\longrightarrow$}}$}\,}

\newcommand*{\clga}{\ensuremath{\mathsf{G}_{\mathsf{A}_1}}} 
\newcommand*{\clc}{\ensuremath{\mathsf{C}}} 
\newcommand*{\cld}{\ensuremath{\mathsf{D}}} 
\newcommand*{\cle}{\ensuremath{\mathsf{E}}} 
\newcommand*{\cli}{\ensuremath{\mathrm{I}}} 
\newcommand*{\fsc}{{}^{\scriptscriptstyle\mathsf{C}}\!} 
\newcommand*{\fsd}{{}^{\scriptscriptstyle\mathsf{D}}\!} 
\newcommand*{\fse}{{}^{\scriptscriptstyle\mathsf{E}}\!} 
\newcommand*{\fsi}{{}^{\scriptscriptstyle\mathrm{I}}\!} 
\newcommand*{\clcg}{\ensuremath{\widetilde{\mathsf{C}}}} 
\newcommand*{\cldg}{\ensuremath{\widetilde{\mathsf{D}}}} 
\newcommand*{\cleg}{\ensuremath{\widetilde{\mathsf{E}}}} 
\newcommand*{\trg}{^{\scriptscriptstyle\triangleright}} 
\newcommand*{\invcmu}{i_{{\scriptscriptstyle\mathsf{C}}_{\!\scriptstyle\mu}}} 
\newcommand*{\bin}{\mathop{\mathrm{bin}}\nolimits} 
\newcommand*{\num}{\mathop{\mathrm{num}}\nolimits} 
\newcommand*{\lcm}{\mathop{\mathrm{lcm}}\nolimits} 
\newcommand*{\bbc}{\ensuremath{\mathbb{C}}} 
\newcommand*{\bbz}{\ensuremath{\mathbb{Z}}} 
\newcommand*{\bbzt}{\ensuremath{\mathbb{Z}\trg}} 

\newcommand*{\cla}{\ensuremath{\mathsf{A}}} 
\newcommand*{\cll}{\ensuremath{\mathsf{L}}} 
\newcommand*{\lmove}{{\scriptstyle{-}1}} 
\newcommand*{\rmove}{{\scriptstyle{+1}}} 
\newcommand*{\undf}{\mathsf{undefined}} 
\newcommand*{\dfa}{\mbox{\normalfont\scshape dfa}}
\newcommand*{\dfas}{\mbox{\normalfont\scshape dfa}s}
\newcommand*{\ow}{\mbox{\normalfont\scshape 1}} 
\newcommand*{\tw}{\mbox{\normalfont\scshape 2}} 

\newcommand*{\iit}[1]{\emph{\/(#1)\/}} 
\newcommand*{\lbr}{\linebreak[0]} 
\newcommand*{\krn}{\ensuremath{\!}} 
\providecommand*{\softl}{l\kern-0.18em\raise0.1ex\hbox{'}\kern-0.04em} 

\newcommand*{\bsection}[2]{\section[#1]{#1\label{#2}}\ignorespaces}

\newtheorem{theorem}{Theorem}
\newcommand*{\btheorem}[1]{\begin{theorem}\label{#1}\ignorespaces}
\newcommand*{\bTheorem}[2]{\begin{theorem}[{#2}]\label{#1}\ignorespaces}
\newcommand*{\etheorem}{\unskip\end{theorem}}

\newtheorem{definition}[theorem]{Definition}
\newcommand*{\bdefinition}[1]{\begin{definition}\label{#1}\upshape\ignorespaces}
\newcommand*{\edefinition}{\unskip\end{definition}}

\newtheorem{lemma}[theorem]{Lemma}
\newcommand*{\blemma}[1]{\begin{lemma}\label{#1}\ignorespaces}
\newcommand*{\elemma}{\unskip\end{lemma}}

\newtheorem{corollary}[theorem]{Corollary}
\newcommand*{\bcorollary}[1]{\begin{corollary}\label{#1}\ignorespaces}
\newcommand*{\ecorollary}{\unskip\end{corollary}}

\newcommand*{\bequations}{\begin{equation}\begin{array}{rcll}}
\newcommand*{\eequations}[1]{\end{array}\label{#1}\end{equation}\ignorespaces}

\newcommand*{\bdisplay}{\[\begin{array}{rcll}}
\newcommand*{\edisplay}{\end{array}\]}

\newcommand*{\benumerate}{\begin{enumerate}}
\newcommand*{\eenumerate}{\unskip\end{enumerate}}
\newcommand*{\eitem}[1]{\item\label{#1}\ignorespaces}

\newcommand*{\bitemize}{\begin{itemize}}
\newcommand*{\eitemize}{\unskip\end{itemize}}

\newcommand*{\bparagraph}[1]{\smallbreak\noindent
  \textbf{{\mathversion{bold}#1.}}\hspace{\labelsep}\ignorespaces}

\providecommand*{\qed}{\mbox{}\nolinebreak\hfill~\raisebox{0.77ex}[0ex]{\framebox[1ex][l]{}}}
\newenvironment{proof}{\noindent\emph{Proof\/}:\hspace{\labelsep}\ignorespaces}{\medbreak}
\newcommand*{\bproof}{\begin{proof}}
\newcommand*{\eproof}{\qed\end{proof}}

\newcommand*{\bfigure}{\unskip\begin{figure}[t]\footnotesize\centering}
\newcommand{\efigure}[2]{\unskip\vspace{-0.2\baselineskip}%
  \caption[x]{#1\label{#2}}\end{figure}\ignorespaces}

\title{Unary Versus Binary Two-Way Automata}%
\author{Viliam Geffert\thanks{Supported by the Slovak Research and
    Development Agency under contract no.~APVV-24-0103\@.}%
  \institute{Department of Computer Science,
    P.\,J.\,\v{S}af\'{a}rik University, Slovakia}%
  \email{viliam.geffert@upjs.sk}%
\and
  Vincent Hlav\'{a}\v{c} \qquad\qquad Rastislav Kr\'{a}lovi\v{c}%
  \institute{Department of Computer Science,
    Comenius University, Slovakia}%
  \email{vincent.hlavac@fmph.uniba.sk\quad\qquad rastislav.kralovic@fmph.uniba.sk}%
}%
\def\titlerunning{Unary Versus Binary Two-Way Automata}%
\def\authorrunning{V. Geffert, V. Hlav\'{a}\v{c} \& R. Kr\'{a}lovi\v{c}}%
\begin{document}
\maketitle
\begin{abstract}%
If \cll~is a unary language, then its \emph{binary coded version}
$\bin\cll$ is a binary language containing all binary strings
representing any $0^x\in \cll$\@. It it known that if a unary
language \cll\ is regular and can be recognized by a minimal one-way
deterministic finite automaton (\ow\dfa) with $n$~states, then its
binary coded version is also regular and can be recognized by a
\ow\dfa\ with at most $n$~states, but at least $1{+}\log n$ states.

Here we shall present related results for two-way automata
(\tw\dfas)\@. First, we shall show that each unary \tw\dfa\
$\cla_1$ with $n$~states can be converted to a \tw\dfa~$\cla_2$
recognizing $\bin\cll(\cla_1)$ with at most $O(n{\cdot}\log n)$
states. If $\cla_1$~is minimal and uses only loops of odd lengths,
$\cla_2$~will use at most $2n{+}2$ states, but it must use at
least $n$~states. For each $n\ge 7$, we shall also present a unary
witness language for which a minimal \tw\dfa\ uses exactly
$n$~states, but any minimal \tw\dfa\ recognizing its binary coded
version uses at least $n$~states, but less than $n{+}\log n$
states.\\
\mbox{}\\
\textbf{Keywords:}\hspace{\labelsep}%
  finite automata, unary regular languages, state complexity
\end{abstract}%

\bsection{Introduction}{s:intro}
If \cll~is a unary language, then its
\emph{binary coded version} $\bin\cll$ is a binary language
containing all binary strings representing any $0^x\in \cll$\@.
Unary (tally) languages play an important role as languages with a
very low information content and many of their properties are
different {}from the general or binary
case~\cite{BMP94b,Ge19,HIS85}\@.

In the case of unary regular languages, we have substantial
differences in state complexity of several operations. As an
example, removing nondeterminism in a one-way nondeterministic
finite automaton with $n$~states may increase the number of states
up to~$2^n$~\cite{RS59,Lu66,Mo71,SWY04}, while the corresponding
cost for automata with unary input alphabet is only
$e^{(1+o(1))\cdot\sqrt{n\cdot\ln n}}$~\cite{Ch86,Ge07,MP01}\@.
There are other differences, for example, if a unary language is
not regular, then it is not context free~\cite{GR62}\@.

\smallbreak
Quite recently~\cite{Ge23,GPS25}, it was shown that if a unary
language \cll\ is recognized by a one-way deterministic finite
automaton (\ow\dfa) $\cla_1$ with $n$~states, then its binary
coded version can be recognized by a \ow\dfa\ $\cla_2$ with at most
$n$~states and, if $\cla_1$~is minimal, then $\cla_2$~must use at
least $1{+}\ceil{\log n}$ states. The gap between unary and binary
versions can indeed be exponential; there are witness languages
matching this gap, for each $n\ge 1$\@.

The conversion in the opposite direction does not work in general:
there exist binary regular languages for which the unary versions
are not regular. As an example, consider the unary language
$\cll= \set{0^{2^k}\st k\ge 0}$\@.
This language is not even context-free, which can be proved
easily, by the use of the Pumping Lemma~\cite[Sect.~7.2]{HMU01}\@.
Nevertheless, its binary coded version is regular: a number is a
power of two if and only if its binary representation contains
exactly one symbol~\quo{$1$} (we allow leading zeros), which can
be tested by a \ow\dfa\ with only $3$~states. However, the set of
powers of two is regular only if the base used for representing
numbers is itself a power of two, but not regular in any other
bases. In general, a set of numbers is regular in all bases only
if its unary coded version is regular, otherwise it can be regular
only in bases which are powers of some fixed positive
integer~\cite{Co69}\@.

In spite of that it is decidable~\cite{Ge25}, for a given
binary \ow\dfa~$\cla_2$, whether there exists a unary \ow\dfa\
$\cla_1$ such that $\bin\cll(\cla_1)= \cll(\cla_2)$\@.
The decision is computed in time that is polynomial in the
number of states of~$\cla_2$\@.
If the answer is positive, the algorithm constructs~$\cla_1$\@.
Interestingly, such decision
becomes easier if we use the convention to write
number representations with the least significant bit first: the
problem of whether there exists a unary \ow\dfa\ $\cla_1$ such
that $\bin\cll(\cla_1)= \cll(\cla_2)\usr{R}$ can be decided in
linear time~\cite{Ma19}\@.

\smallbreak
This paper presents related results for two-way deterministic
finite automata (\tw\dfas)\@. First, we shall show that each unary
\tw\dfa\ $\cla_1$ with $n$~states can be converted to a binary
\tw\dfa~$\cla_2$ recognizing $\bin\cll(\cla_1)$ with at most
$O(n{\cdot}\log n)$ states.

Moreover, if $\cla_1$~does not use loops of even lengths, $2n{+}2$
states are sufficient. If, in addition, $\cla_1$~is minimal, then
$n$~states are necessary. This follows {}from a conversion in the
opposite way: if $\cla_2$~is a binary \tw\dfa\ with $n$~states
recognizing a binary coded version of a unary language~\cll\ with no
tail and an odd period, it can be converted to a a unary
\tw\dfa~$\cla_1$ recognizing~\cll\ with at most $n$~states.
Next, for each $n\ge 7$, we shall present a unary witness
language~$\cll_n$ for which a minimal \tw\dfa\ uses exactly
$n$~states such that any minimal binary \tw\dfa\ recognizing
$\bin\cll_n$ must use at least $n$~states, but less than
$n+\log n -1$ states.

\bsection{Preliminaries}{s:prel}
Here we shall briefly fix some basic definitions and notation on
finite state automata. For more details, we refer the reader
to~\cite{HMU01}, or any other standard textbook.

\smallbreak
A~\emph{two-way deterministic finite state automaton} (\tw\dfa,
for short) is defined as a quintuple
$\cla =(Q,\Sigma,\lbr \delta,\lbr q\xsr{I},F)$, where $Q$~denotes a
finite set of states, $\Sigma$~a~finite set of input symbols,
$q\xsr{I}\in Q$ an~initial state, and $F\subseteq Q$ a~set of final
(accepting) states. The states in the set $Q\setminus F$ will be
called rejecting. Finally,
$\delta: Q\times(\Sigma\cup\set{\vdash,\dashv})\rightarrow
  Q\times\set{\lmove,\rmove}$
is a~\emph{partially defined} transition function. Here $\vdash$
and~$\dashv$ denote the left and right input tape endmarkers,
respectively, such that $\Sigma\cap\set{\vdash,\dashv}= \emset$,
and $\lmove,\rmove$ represent the directions of the input head
movement (to the left or right, respectively)\@.

\cla~operates on~${\vdash}w{\dashv}$, an input word
$w\in \Sigma\str$ surrounded on the tape by the two endmarkers,
starting in~$q\xsr{I}$ with the input head positioned on~${\vdash}$\@.
A~transition $\delta(q,a)= (q'\krn,d)$ is interpreted as follows:
if \cla~is in the state~$q$ and the head reads the symbol~$a$, the
state switches to~$q'$ and the head moves one position to the left
or right, depending on the value $d\in \set{\lmove,\rmove}$\@. The
automaton cannot leave~${\vdash}w{\dashv}$, that is,
$\delta(q,\vdash)\ne (q'\krn,\lmove)$ and
$\delta(q,\dashv)\ne (q'\krn,\rmove)$, for each $q,q'\in Q$\@. If
$\delta(q,a)$ is not defined, that is, if $\delta(q,a)= \undf$,
the machine halts. The input is accepted if, after a finite number
of steps, \cla~halts in an accepting state $q\in F$\@. If
\cla~halts in $q\notin F$ or if it does not halt at all (getting
into an infinite cycle), the input is rejected. The set of all
inputs accepted by~\cla\ is the
\emph{language}~$\cll(\cla)$\@.

\smallbreak
The automaton is called \emph{sweeping}, if the direction of the
input head movement changes only on the endmarkers.

The automaton is \emph{one-way} (\ow\dfa, for short), if it never
moves the head back to the left, that is, if
$\delta: Q\times(\Sigma\cup\set{\vdash,\dashv})\rightarrow
  Q\times \set{\rmove}$\@.
According to this, a~\ow\dfa\ has also the input delimited by the
endmarkers, which differs {}from the standard definition in the
literature~\cite{HMU01}\@. However,
conversion to the standard model does not
cost more than two new states.

Similarly, the definition of acceptance for two-way automata may
differ in the literature: a~\tw\dfa\ is sometimes required to
accept at the right endmarker, sometimes to move its head to the
right of~$\dashv$, or to leave ${\vdash}w{\dashv}$ in either
direction. (Cf., e.g.,
\cite{GO21,Ka05m,Ka09,KO11}\@.) None of
these peculiarities matter, as long as we are interested in the
asymptotic number of states\,---\,such differences can be fixed by
adding a constant number of new states.

Two automata are
\emph{equivalent}, if they recognize the same language. A~\ow\dfa\
(\tw\dfa) \cla~is \emph{minimal}, if no \ow\dfa\ (\tw\dfa,
respectively) with fewer states is equivalent to~\cla\@.

\smallbreak
The binary representation of natural
numbers is defined in a standard way, by the use of functions
$\num: \set{0,1}\str\krn\rightarrow \bbn$ and
$\bin: \bbn\rightarrow \set{0,1}\str$\krn, as follows:%
\footnote{To distinguish between the standard multiplication of
  integers and a concatenation of strings in formulas with mixed
  contents, strings are sometimes enclosed in quotation marks\@.}
\bdisplay
  {\num}\mathquo{\eps} &=& 0 \,,\\
  {\num}\mathquo{w{\cdot}b} &=& ({\num}\mathquo{w}){\cdot}2 +b \,,
    &\mbox{ for each $w\in \set{0,1}\str$ and $b\in \set{0,1}$},\\[0.5ex]
  \bin x &=& \makebox[0em][l]{the shortest $w\in \set{0,1}\str$
    such that ${\num}\mathquo{w}= x$\@.}
\edisplay

Next, let $\cll_1\subseteq \set{0}\str$ and
$\cll_2\subseteq \set{0,1}\str$\krn\@. Then
\bdisplay
  \bin\cll_1= \set{w\in \set{0,1}\str\st 0^{\num w}\in \cll_1} \,,
    & \ &
  \num\cll_2= \set{0^{\num w}\st w\in \cll_2} \,.
\edisplay

\smallbreak
We finish this section by an auxiliary lemma, required later
(without a proof here, due to space constraints)\@.
It is known that each $n\ge 7$, except for $n= 9$,
can be partitioned to a sum of distinct odd primes~\cite{Dr72}\@.
However, \cite{Dr72}~does not pay any attention to the
\emph{number of primes} forming the sum, which we would like to
have as small as possible.

\blemma{l:primes}
Each $n\ge 7$, except for $n= 9$, can be expressed as a sum of
less than $\log n$ odd primes, all different. Moreover, if
$n\ge 23$, none of these primes is larger than $n{-}11$\@.
\elemma

\bsection{Some Facts About One-Way and Unary Automata}{s:old}
Here we shall present some results about one-way automata, mainly
{}from~\cite{Ge23,Ge25,GPS25}, as well as about unary two-way
automata, mainly {}from~\cite{KO11,KO12}, required later.

\smallbreak
A unary one-way automaton $\cla_1$ is very simple.
The state set of~$\cla_1$ is
$Q_1= \bbzt_{\sigma}\cup\bbz_{\lambda}$, where
$\bbzt_{\sigma}= \set{0\trg\!,1\trg\!,\ldots,(\sigma{-}1)\trg}$ is
an \emph{initial segment}%
\footnote{Both in $\bbzt_{\sigma}$ and in~$\bbz_{\lambda}$, the
  states are associated with integers. To distinguish states in
  the initial segment {}from those in the loop, the former are
  labeled by triangles while the latter are not\@.}
and $\bbz_{\lambda}= \set{0,1,\ldots,\lambda{-}1}$ a~\emph{loop}\@.
In the initial segment, $\cla_1$~counts
the length of the input up to~$\sigma{-}1$ after which, in the
loop, it counts modulo~$\lambda$\@:
$\delta_1(j\trg\!,0)= (j{+}1)\trg$ for $j\le \sigma{-}2$, but
$\delta_1((\sigma{-}1)\trg\!,0)= \sigma\mmod\lambda$, and
$\delta_1(i,0)= (i{+}1)\mmod\lambda$\@. If $\sigma= 0$, there is
no initial segment and $\bbzt_{\sigma}= \emset$\@. Depending on
whether $\sigma> 0$ or $\sigma= 0$, the initial state is
$q\xsr{I,1}= 0\trg$ or $q\xsr{I,1}= 0$\@.

It is obvious that $\cll(\cla_1)$ has a~\emph{period}
$\lambda\ge 1$ and a~\emph{tail} $\sigma\ge 0$, that is,
$0^N\in \cll(\cla_1)$ if and only if
$0^{N+\lambda}\in \cll(\cla_1)$, for each $N\ge \sigma$\@.

The unary \ow\dfa\ $\cla_1$ can be converted to a binary~$\cla_2$,
such that $\cll(\cla_2)= \bin\cll(\cla_1)$\@. The \quo{basic}
conversion is quite straightforward, without increasing the number
of states:

\bTheorem{t:owbas}{\cite[Thm.~2]{GPS25}}
If a unary language \cll\ is recognized by a \ow\dfa\ $\cla_1$ using
$n$ states, namely, an initial segment~$\bbzt_{\sigma}$ with
$\sigma$~states and a loop~$\bbz_{\lambda}$ with $\lambda$~states,
then $\bin\cll$ is also regular and can be recognized by a
\ow\dfa~$\cla_2$ using $n$~states, namely, a \emph{preamble} with
$\sigma$~states (a~pre-periodic component, simulating the original
initial segment) and a \emph{kernel} with $\lambda$~states
(a~periodic component, simulating the original loop)\@.
\etheorem
This \quo{basic} binary \ow\dfa\ $\cla_2$ uses the same set of
states $\bbzt_{\sigma}\cup\bbz_{\lambda}$ as well as the same
initial and accepting states, but with different transitions,
for each $b\in \set{0,1}$:
\bequations
  \delta_2(j\trg\!,b) &=& \left\{\begin{array}{ll}
    (j{\cdot}2{+}b)\trg , &\mbox{ if $j{\cdot}2{+}b< \sigma$}, \\
    (j{\cdot}2{+}b)\mmod\lambda \,, &\mbox{ if $j{\cdot}2{+}b\ge \sigma$},
  \end{array}\right. \\[2.0ex]
  \delta_2(i,b) &=& (i{\cdot}2{+}b)\mmod\lambda \,.
\eequations{e:owbas}
Thus, instead of counting symbols along a unary input,
$\cla_2$~computes the numerical value ${\num}\mathquo{w}$ for the
prefix~$w$ which, so far, has been read {}from the binary
input. When ${\num}\mathquo{w}$ exceeds~\mbox{$\sigma{-}1$},
\,$\cla_2$~starts to compute this value modulo~$\lambda$\@.
Despite the fact that $\cla_2$ reflects the structure of the
original minimal unary~$\cla_1$, it is usually far {}from being
minimal.

It turns out that a partial factorization of the loop
length~$\lambda$ plays an important role here; let us express
it in the form
$\lambda= \mu\cdot 2^{\ell}$,
where $\mu\ge 1$ is odd and $\ell\ge 0$\@.
Using this, the cost of conversion {}from
unary to binary one-way automata can be summarized as follows:

\btheorem{t:owmin}
Let $\cla_1$ be a minimal unary \ow\dfa\ using $n$~states, namely,
$\sigma$~states in the initial segment and
$\lambda= \mu{\cdot}2^{\ell}$ states in the loop, where $\mu$~is
odd and $\sigma+\mu{\cdot}2^{\ell}= n$\@. Then the
minimal \ow\dfa\ recognizing $\bin\cll(\cla_1)$, uses
  \iit{a}~at most $\mu{\cdot}2^{\ell}$ but at least
    $\mu{+}\ell$ states in the kernel (for simulation of the
    original loop),
  \iit{b}~at most $\sigma$ but at least
    $\max\set{1,1{+}\ceil{\log\sigma}{-}\ell}$ states outside the
    kernel (for simulation of the original initial segment),
    except for $\sigma= 0$, with no states outside the
    kernel,
  \iit{c}~at most~$n$ but at least $1{+}\ceil{\log n}$
    states in total.
\etheorem
It should be pointed out that all these bounds are exact and
cannot be improved. That is, there exist witness languages
matching all upper and lower bounds presented in the above
theorem, for any given $\sigma,\mu,\ell$\@. For more details,
see~\cite{Ge23,GPS25}\@.

\smallbreak
The opposite conversion
is not always possible, as already presented in
Section~\ref{s:intro}\@. Nevertheless, the following has been
obtained for such conversion~\cite{Ge25}:

\bTheorem{t:decd}{\cite[Cor.~14, Alg.~1]{Ge25}}
It is decidable, for a given minimal binary \ow\dfa~$\cla_2$, whether
there exists a corresponding unary \ow\dfa\ $\cla_1$ such that
$\bin\cll(\cla_1)= \cll(\cla_2)$\@. Moreover, if the answer is
positive, the algorithm constructs~$\cla_1$\@.
\etheorem
The decision itself, as well as the lengths of the initial segment and
the loop in~$\cla_1$ (if~$\cla_1$ does exist) are computed in time
that is polynomial in the number of states of~$\cla_2$\@. After
that, the resulting unary $\cla_1$ is produced in time that is
polynomial in the number of states of~$\cla_1$\@. (The gap between
the sizes of $\cla_2$ and~$\cla_1$ may be exponential\@.)

\smallbreak
Before passing further, we need some facts about unary
two-way automata, namely, the result by Kunc and
Okhotin~\cite[Thm.~2]{KO11} stating that a unary \tw\dfa\ $\cla_1$
with $n$~states can be transformed to an equivalent \tw\dfa\ with
$n{+}1$ states that is sweeping.%
\footnote{Without proof, Chrobak stated that this can be done
  without increasing the number of states already in~1986,
  in~\cite{Ch86}\@.}
The original proof in~\cite{KO11} is based on a two-way analogue
of transformation semigroups. Since we shall need to see some
details about the behavior of such automata, like properties of
\emph{cardinal states} and \emph{inner graphs}, we are going to
provide an alternative direct construction.

\bdefinition{d:graph}
Given~a unary \tw\dfa\ $\cla_1$, the \emph{inner graph
of~$\cla_1$} is a directed graph~\clga\ in which vertices
represent the states of~$\cla_1$ and labeled edges represent
transitions not reading any of the endmarkers, i.e., an edge
$q\trs{d}q'$ represents a~transition $\delta(q,0)= (q'\krn,d)$\@.

A~set $\clc= \set{p_0,\ldots,p_{k-1}}$ connected in this graph by
edges $p_0\trs{d_0} p_1\trs{d_1} \cdots p_{k-1}\trs{d_{k-1}} p_0$,
for some $d_0,\ldots,d_{k-1}\in \set{\lmove,\rmove}$, is called
a \emph{loop, of length} $\fsc\lambda= \sum_{i=0}^{k-1} d_i$\@.
\,\clc~is a~\emph{left-to-right} loop, if $\fsc\lambda> 0$,
a~\emph{right-to-left} loop, if $\fsc\lambda< 0$, and
a~\emph{not-moving} loop, if $\fsc\lambda= 0$\@.

A~set $\cli= \set{p_0,\ldots,p_{k-1}}$ connected by
$p_0\trs{d_0} p_1\trs{d_1} \cdots p_{k-1}\trs{d_{k-1}} p$,
with $p_0,\ldots,p_{k-1}$ outside any loop, $p_0$~equal to the
initial state, and $p$~inside a loop, is an
\emph{initial segment}\@.

The unary \tw\dfa\ $\cla_1$ is in \emph{cyclic form}, if each
vertex of~\clga\ belongs to some loop or to the initial segment.
(If the initial state belongs to some loop, there is no initial
segment\@.)
\edefinition

Note that, rather than~$\abs{C}$, the~number of states in the given loop~\clc,
its length~$\fsc\lambda$ represents the distance traveled along
the input tape in the course of one iteration. This value is
negative for loops moving to the left. If the automaton is
sweeping, $\abs{C}= \abs{\fsc\lambda}$\@.

\bdefinition{d:card}
Given a unary \tw\dfa\ $\cla_1$, let~\bbc\ denote the set of all
loops in the inner graph of~$\cla_1$\@. In each loop
$\clc\in \bbc$ with $\fsc\lambda\ne 0$, fix the set of
\emph{cardinal states} $\clc'\subseteq \clc$, as follows:
\bfigure
\unitlength.35mm%
\linethickness{0.4pt}%
\ifx\plotpoint\undefined\newsavebox{\plotpoint}\fi
\begin{picture}(244,113)(0,0)
\put(100,66){\circle{8}}
\put(128,66){\circle{8}}
\put(60,66){\circle{8}}
\put(184,10){\circle{8}}
\put(4,10){\circle{8}}
\put(212,10){\circle{8}}
\put(32,10){\circle{8}}
\put(60,10){\circle{8}}
\put(4,94){\circle{8}}
\put(32,94){\circle{8}}
\put(60,94){\circle{8}}
\put(4,38){\circle{8}}
\put(128,38){\circle{8}}
\put(156,38){\circle{8}}
\put(60,38){\circle{8}}
\put(18,92){\makebox(0,0)[ct]{$\scriptstyle{+}1$}}
\put(18,12){\makebox(0,0)[b]{$\scriptstyle{+}1$}}
\put(46,92){\makebox(0,0)[ct]{$\scriptstyle{+}1$}}
\put(46,12){\makebox(0,0)[b]{$\scriptstyle{+}1$}}
\put(59,80){\makebox(0,0)[r]{$\scriptstyle{-}1$}}
\put(59,52){\makebox(0,0)[r]{$\scriptstyle{-}1$}}
\put(59,25){\makebox(0,0)[r]{$\scriptstyle{+}1$}}
\put(6,25){\makebox(0,0)[lc]{$\scriptstyle{+}1$}}
\put(6,52){\makebox(0,0)[lc]{$\scriptstyle{-}1$}}
\put(130,111){\makebox(0,0)[lc]{$r_1$}}
\put(158,111){\makebox(0,0)[lc]{$r_2$}}
\put(183,56){\makebox(0,0)[rc]{$r_3$}}
\put(211,56){\makebox(0,0)[rc]{$r_4$}}
\put(239,28){\makebox(0,0)[rc]{$r_5$}}
\put(96,70){\makebox(0,0)[rb]{$V$}}
\put(124,98){\makebox(0,0)[rb]{$S$}}
\put(124,70){\makebox(0,0)[rb]{$U$}}
\put(151,98){\makebox(0,0)[rb]{$T$}}
\put(160,42){\makebox(0,0)[lb]{$X$}}
\put(180,6){\makebox(0,0)[rt]{$R$}}
\put(208,6){\makebox(0,0)[rt]{$S$}}
\put(234,6){\makebox(0,0)[rt]{$T$}}
\put(132,42){\makebox(0,0)[lb]{$W$}}
\put(189,42){\makebox(0,0)[lb]{$Y$}}
\put(217,42){\makebox(0,0)[lb]{$Z$}}
\put(0,98){\makebox(0,0)[rb]{$R$}}
\put(28,98){\makebox(0,0)[rb]{$S$}}
\put(56,98){\makebox(0,0)[rb]{$T$}}
\put(64,70){\makebox(0,0)[lb]{$U$}}
\put(64,42){\makebox(0,0)[lb]{$V$}}
\put(0,42){\makebox(0,0)[rb]{$Z$}}
\put(64,14){\makebox(0,0)[lb]{$W$}}
\put(28,6){\makebox(0,0)[rt]{$X$}}
\put(0,14){\makebox(0,0)[rb]{$Y$}}
\put(8,94){\vector(1,0){20}}
\put(132,94){\vector(1,0){20}}
\put(36,94){\vector(1,0){20}}
\put(124,66){\vector(-1,0){20}}
\put(160,38){\vector(1,0){20}}
\put(188,38){\vector(1,0){20}}
\put(132,38){\vector(1,0){20}}
\put(188,10){\vector(1,0){20}}
\put(4,14){\vector(0,1){20}}
\put(60,90){\vector(0,-1){20}}
\put(60,62){\vector(0,-1){20}}
\put(60,34){\vector(0,-1){20}}
\put(216,10){\vector(1,0){20}}
\put(28,10){\vector(-1,0){20}}
\put(56,10){\vector(-1,0){20}}
\put(103,63){\vector(1,-1){22}}
\put(4,42){\vector(0,1){48}}
\put(180,34){\framebox(8,8)[cc]{$\bullet$}}
\put(152,90){\framebox(8,8)[cc]{$\bullet$}}
\put(124,90){\framebox(8,8)[cc]{}}
\put(208,34){\framebox(8,8)[cc]{$\bullet$}}
\put(236,6){\framebox(8,8)[cc]{$\bullet$}}
\put(155.766,101.766){\line(0,1){1.75}}
\put(155.766,105.266){\line(0,1){1.75}}
\put(155.766,108.766){\line(0,1){1.75}}
\put(155.766,112.266){\line(0,1){1.75}}
\put(127.766,101.766){\line(0,1){1.75}}
\put(127.766,105.266){\line(0,1){1.75}}
\put(127.766,108.766){\line(0,1){1.75}}
\put(127.766,112.266){\line(0,1){1.75}}
\put(211.766,45.766){\line(0,1){1.75}}
\put(211.766,49.266){\line(0,1){1.75}}
\put(211.766,52.766){\line(0,1){1.75}}
\put(211.766,56.266){\line(0,1){1.75}}
\put(239.766,17.766){\line(0,1){1.75}}
\put(239.766,21.266){\line(0,1){1.75}}
\put(239.766,24.766){\line(0,1){1.75}}
\put(239.766,28.266){\line(0,1){1.75}}
\put(183.766,45.766){\line(0,1){1.75}}
\put(183.766,49.266){\line(0,1){1.75}}
\put(183.766,52.766){\line(0,1){1.75}}
\put(183.766,56.266){\line(0,1){1.75}}
\put(152,90){\vector(-1,-1){21}}
\put(208,34){\vector(-1,-1){21}}
\end{picture}%
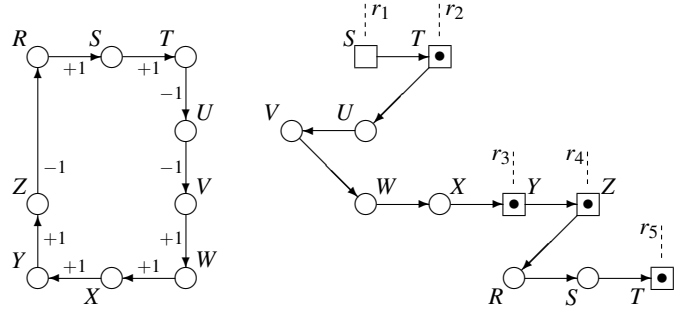
\efigure{Fixing cardinal states for a left-to-right loop. The
  given loop~\clc\ in the inner graph of~$\cla_1$ is displayed on
  the left, the corresponding movement of~$\cla_1$ along the input
  tape is shown on the right. By choosing $r_1= S$, the sequence
  $r_1,r_2,\ldots$ is $S,T,Y,Z,T$\,---\,along the trajectory of
  movements, these states are displayed as squares and pointed to
  by vertical dashed lines. The first repeated state is~$T$ (i.e.,
  $r_5= r_2$), which gives the set of cardinal states
  $\clc'=\set{T,Y,Z}$\,---\,labeled by bullets. The corresponding
  left-to-right loop in a sweeping \tw\dfa\ will thus be
  $T\trs{{+}1} Y\trs{{+}1} Z\trs{{+}1} T$\@.}{f:card}

If \clc~is a left-to-right loop, fix one state in~\clc\
as~$r_1$\,---\,the choice is quite arbitrary. After fixing
$r_1,\ldots,r_j$, fix $r_{j+1}$ as the state in which $\cla_1$
gets to the right of~$r_j$ for the first time, along the path
starting in~$r_j$\@. (For an illustrating example, see
Figure~\ref{f:card}\@.) Such state must exist, since
\clc~traverses to the right. This is repeated for
$j= 1,2,3,\ldots$ until, for some~$j$, the sequence
$r_1,\ldots,r_j$ enumerates the same state twice, that is, until
$r_j= r_i$ for some $i< j$\@. After that, denote
$r_i,r_{i+1}\ldots,r_{j-1}$ (excluding~$r_j$) by
$\fsc q_0,\fsc q_1,\ldots,$ respectively, and declare
$\clc'= \set{\fsc q_0,\fsc q_1,\ldots}$ as the set of
\emph{cardinal states} in the loop~\clc\@.

If \clc~is a right-to-left loop, the procedure is symmetrical:
after fixing $r_1,\ldots,r_j$, fix $r_{j+1}$ such that
$\cla_1$~gets \emph{to the left} of~$r_j$ for the first time.
After obtaining $r_j= r_i$ for some $i< j$, denote
$r_i,r_{i+1}\ldots,r_{j-1}$ by $\fsc q_0,\fsc q_1,\ldots$ and
declare $\clc'= \set{\fsc q_0,\fsc q_1,\ldots}$ as the set of
\emph{cardinal states} in~\clc\@.
\edefinition

Clearly, if \clc~is a left-to-right loop, each cardinal state
$q\in \clc'\subseteq \clc$ is visited along~\clc\ by a transition
\emph{coming {}from the left}, i.e., by $\delta(p,0)= (q,\rmove)$,
for some $p\in \clc$\@. This does not exclude visiting
$q\in \clc'$ {}from the right, but only {}from states outside the
loop~\clc\@. Next, since one iteration of~\clc\ travels exactly
$\fsc\lambda$~input positions, visits all states
in~$\clc'$, but none of them is visited twice,
$\clc'= \set{\fsc q_0,\fsc q_1,\ldots}$ contains exactly
$\fsc\lambda$ states. Thus, starting {}from a state
$\fsc q_e\in \clc'$\krn, the head gets for the first time
$h\ge 0$ positions to the right in the state $\fsc q_k$, where
$k= (e{+}h)\mmod\fsc\lambda$, provided that the head does not hit
an endmarker in the meantime.
Symmetrically, if \clc~is a right-to-left loop, each cardinal
state $q\in \clc'$ is visited by a transition coming along~\clc\
\emph{{}from the right} and $\clc'= \set{\fsc q_0,\fsc q_1,\ldots}$
contains exactly ${-}(\fsc\lambda)> 0$ states\@.

\blemma{l:card}
Let $\cla_1$ be a unary \tw\dfa\ with $n$~states, with loops of
lengths $\lambda_1,\ldots,\lambda_r$ in the inner graph, and let
$n_0= n-\sum_{i=1}^{r} \abs{\lambda_i}$\@.

If~$\cla_1$, starting in any state at the left endmarker, moves
the head $n_0{+}2$ positions away, it gets to this position in a
cardinal state of a left-to-right loop~\clc, after which it
traverses across the input and hits the right endmarker in a
(possibly different) cardinal state of~\clc\@.

The corresponding symmetrical statement holds for computations
that move $n_0{+}2$ positions away {}from the right endmarker,
traversing the entire input {}from right to left.
\elemma

\bTheorem{t:swwep}{\cite{KO11}}
For each unary \tw\dfa\ $\cla_1$ with $n$~states, there exists an
equivalent \tw\dfa\ $\cla'_1$ with $n{+}1$ states that is
sweeping, in cyclic form.

More precisely, if $\cla_1$~uses loops of lengths
$\lambda_1,\ldots,\lambda_r$ in the inner graph, $\cla'_1$~uses
$\sum_{i=1}^{r} \abs{\lambda_i}$ states in the respective loops of
the same lengths (which eliminates loops of length zero) and at
most $n{+}1- \sum_{i=1}^{r}\abs{\lambda_i}$ states in an initial
segment.
\etheorem
\bproof
Let~\bbc\ be the set of all loops in the inner graph of~$\cla_1$
and let $n_0= n-\sum_{\clc\in \bbc} \abs{\fsc\lambda}$\@.
As an initial assumption,
suppose the following:
\benumerate
  \eitem{t:swwep.ass} Starting {}from the initial state at the left
    endmarker, $\cla_1$~visits the input tape position $n_0{+}2$
    on any input of length $N\ge n_0{+}1$\@.
\eenumerate
The pathological case of~$\cla_1$ that never moves its head more
than $n_0{+}1$ positions away {}from the left endmarker will be
left to reader, due to space constraints. (Should this be the case, the
language $\cll(\cla_1)$ must be either finite or cofinite\@.)

$\cla'_1$~begins by deciding the membership in~$\cll(\cla_1)$ for
inputs shorter than $n_0{+}1$ by the use of the initial segment
$\cli= \set{\fsi q_0,\fsi q_1,\ldots,\fsi q_{n_0}}$ consisting
of $n_0{+}1$ new states, where $\fsi q_0$~is the initial state,
with the following transitions:
\bitemize
 \item $\delta'(\fsi q_0,\vdash)= (\fsi q_0,\rmove)$,
 \item $\delta'(\fsi q_e,0)= (\fsi q_{e+1},\rmove)$,
    \,for each $e\in \set{0,\ldots,n_0{-}1}$,
 \item $\delta'(\fsi q_e,\dashv)= \undf$,
    \,for each $e\in \set{0,\ldots,n_0}$\@.\\
    Depending on whether $0^e\in \cll(\cla_1)$, the state
    $\fsi q_e$ is made accepting or rejecting.
\eitemize

{}From this moment on, all subsequent transitions can rely on the
fact that the input is of length $N\ge n_0{+}1$, that is, the
endmarkers are at least~$n_0{+}2$ positions away {}from each other.
The next transition $\delta'(\fsi q_{n_0},0)$ requires a special
care:
\bitemize
  \item $\delta'(\fsi q_{n_0},0)= (q,\rmove)$,
    \,where $q$~is the first state in which, starting {}from the
    initial state at the left endmarker, $\cla_1$~gets to the
    position~$n_0{+}2$\@.
\eitemize
By the assumption~(\ref{t:swwep.ass}), such $q$~does exist and, by
Lemma~\ref{l:card}, the state $q$ must be a cardinal state in a
left-to-right loop, that is, $q= \fsc q_e$ for some
$\fsc q_e\in \clc'\subseteq \clc\in \bbc$\@.

\smallbreak
Using Lemma~\ref{l:card} again, each traversal {}from left to
right across the input hits the right endmarker in a cardinal
state of a left-to-right loop. (Symmetrically, a traversal in the
opposite direction ends in a cardinal state of a right-to-left
loop\@.) Once we get into a \quo{proper place in a proper
loop}\krn, we worry only about a cardinal state in which
we hit an endmarker. This allows us to replace the loops
of~$\cla_1$ by $\bigcup_{\clc\in\bbc}\clc'$\krn, the set of all
cardinal states in these loops.%
\footnote{This eliminates all not-moving loops, with
  $\clc'= \emset$\@.}
In~$\cla'_1$, transitions traveling across the input are
simple, counting modulo $\abs{\fsc\lambda}$ while moving the
head in proper direction: for each $\clc\in \bbc$ and each
$\fsc q_e\in \clc'$\krn,
\bitemize
  \item $\delta'(\fsc q_e,0)= (\fsc q_{(e+1)\mmod\abs{\fsc\lambda}},\,d)$,
    \,where $d= \rmove$ or $d= \lmove$,\\
    depending on whether $\fsc\lambda> 0$ or $\fsc\lambda< 0$\@.
\eitemize

It remains to implement switching {}from a left-to-right traversal
across the input to a right-to-left traversal taking place near
the right endmarker and, symmetrically, {}from a right-to-left
traversal to a left-to-right traversal near the left endmarker.
Consider first the latter, that is, $\cla_1$ arrives to~$\vdash$
in a state $\fsd q_j$ belonging to a right-to-left loop
$\cld\in \bbc$\@. This situation is handled as follows:
\bitemize
  \item $\delta'(\fsd q_j,\vdash)= (\fsc q_i,\rmove)$,
      \,where $i= (e{+}1{-}(n_0{+}2))\mmod\fsc\lambda$,\\
    if the path starting {}from~$\fsd q_j$ gets $n_0{+}2$ positions
    away {}from~$\vdash$, the first time in the state~$\fsc q_e$
    belonging to a loop \clc\ of length~$\fsc\lambda$\@.
  \item $\delta'(\fsd q_h,\vdash)= \undf$,
    \,if the path starting {}from~$\fsd q_j$ halts in a state~$q$,
    not moving farther than $n_0{+}1$ positions away
    {}from~$\vdash$\@. Depending on whether $q\in F_1$, the
    state $\fsd q_h$ is made accepting or rejecting.
  \item $\delta'(\fsd q_h,\vdash)= \undf$,
    \,if the path starting {}from~$\fsd q_j$ does not halt in
    $n{\cdot}(n_0{+}2)$ steps,%
    \footnote{Which means that $\cla_1$ does
      not halt at all\@.}
    not moving farther than $n_0{+}1$
    positions away {}from~$\vdash$\@. The state
    $\fsd q_h$ is made rejecting.
\eitemize
This is established for each right-to-left loop $\cld\in \bbc$ and
each $\fsd q_j\in \cld'$\krn\@.

The reasoning behind this is based on
Lemma~\ref{l:card}: if~$\cla_1$, starting {}from~$\fsd q_j$
at~$\vdash$, moves%
\footnote{Before getting to the position $n_0{+}2$, the path
  starting {}from~$\fsd q_j$ may return to~$\vdash$ several times.
  However, we apply Lemma~\ref{l:card} to the segment connecting
  the \emph{last visit} at the endmarker with the
  position~$n_0{+}2$\@.}
to the position $n_0{+}2$, it gets there in a state that is
cardinal, i.e., in some $\fsc q_e$ belonging to a
left-to-right loop~\clc, of length~$\fsc\lambda$\@. After that,
$\cla_1$~iterates~\clc\ until it gets to~$\dashv$ at the
position~$N{+}1$, getting there in~$\fsc q_k$, with
$k= (e{+}(N{+}1){-}(n_0{+}2))\mmod\fsc\lambda$\@. In the
sweeping~$\cla'_1$, the original loop~\clc\ is replaced
by~$\clc'$ composed of cardinal states, but traveling
the same distance~$\fsc\lambda$\@. By entering~$\clc'$ in the
\quo{proper} position, we start iteration right after
leaving~$\vdash$\@. Namely, starting {}from the position~$1$
in~$\fsc q_i$, where $i= (e{+}1{-}(n_0{+}2))\mmod\fsc\lambda$, the
new automaton $\cla'_1$ gets to~$\dashv$ in~$\fsc q_{k'}$, where
$k'= (i+ (N{+}1) -1)\mmod\fsc\lambda=
  ((e{+}1{-}(n_0{+}2))+ (N{+}1) -1)\mmod\fsc\lambda= k$\@.
Thus, $\cla'_1$~gets to~$\dashv$ in the same state as
does~$\cla_1$\@.
The remaining items fix cases in which~$\cla_1$,
after arriving to~$\vdash$, does not move farther than $n_0{+}1$
positions away {}from~$\vdash$ any more.

\smallbreak
Switching {}from left-to-right loops to right-to-left loops at the
right endmarker is symmetrical, just swapping the roles
of~$\vdash,\dashv$, the roles of left/right input head movements,
and computing distances modulo~${-}(\fsc\lambda)$, which we
leave to an interested reader.

Finally, since $\cla'_1$~never visits~$\vdash$ in a state
belonging to a left-to-right loop, or $\dashv$~in a state of a
right-to-left loop, we can leave transitions for these situations
$\undf$\@.

The total number of states in the loops is
$\sum_{\clc\in \bbc} \abs{\fsc\lambda}$ and the length of the
initial segment is bounded by
$\abs{\cli}= n_0{+}1= n{+}1- \sum_{\clc\in \bbc}\abs{\fsc\lambda}$\@.
\eproof

If a finite number of short inputs do not matter, the initial
segment can be removed:

\bcorollary{c:swwep}
For each unary \tw\dfa\ $\cla_1$ with $n$~states, using loops of
lengths $\lambda_1,\ldots,\lambda_r$ in the inner graph, there
exists a \tw\dfa\ $\cla''_1$ with
$\sum_{i=1}^{r}\abs{\lambda_i}\le n$ states that is sweeping, in
cyclic form with no initial segment, using loops of the same
lengths (with eliminated loops of length zero), such that
$\cla''_1$ agrees with~$\cla_1$ on all inputs of length
$N> n- \sum_{i=1}^{r}\abs{\lambda_i}$\@.
\ecorollary
\bproof
Our starting point is the construction of an equivalent $\cla'_1$
{}from Theorem~\ref{t:swwep}, with $\sum_{i=1}^{r}\abs{\lambda_i}$
states in the respective loops and an initial segment.
Recall that $\cla'_1$ begins by passing through an initial segment
of $n_0{+}1$ states, where
$n_0= n- \sum_{i=1}^{r}\abs{\lambda_i}$\@. After that, if the
input is of length at least $n_0{+}1$, \,$\cla'_1$~enters the
position $n_0{+}2$ in some~$\fsc q_e$, a state in a loop
$\clc'= \set{\fsc q_0,\fsc q_1,\ldots,\fsc q_{\fsc\lambda-1}}$
of length $\fsc\lambda> 0$, and starts iteration of this loop.

Thus, by getting\,---\,in an arbitrary way\,---\,to the state
$\fsc q_e$ at the position $n_0{+}2$, the outcome of the
computation does not change, which holds for all inputs of length
at least $n_0{+}1$\@. Therefore, we can start the iteration
of~$\clc'$ earlier, right after leaving~$\vdash$, at the
position~$1$, in the state $\fsc q_h\in \clc'$\krn, where
$h= (e{+}1{-}(n_0{+}2))\mmod\fsc\lambda$\@. Next, by the
construction given in Theorem~\ref{t:swwep}, we see that
$\cla'_1$~never visits the left endmarker in a state belonging to
a left-to-right loop. This leaves
$\delta'(\fsc q_h,\vdash)= \undf$\@. But then we can obtain
$\cla''_1$ {}from~$\cla'_1$ by redefining this transition to
$\delta'(\fsc q_h,\vdash)= (\fsc q_h,\rmove)$ and by making
$\fsc q_h$ a new initial state.

After that, the initial segment
can be removed.
As a result, $\cla''_1$~consists only of loops in~$\cla'_1$, which
gives $\sum_{i=1}^{r}\abs{\lambda_i}\le n$ states. $\cla''_1$~agrees
in acceptance/rejection with~$\cla'_1$ (hence, also with~$\cla_1$)
on all inputs of length
$N\ge n_0{+}1 > n- \sum_{i=1}^{r}\abs{\lambda_i}$\@.
\eproof

\bsection{Binary Two-Way Automata}{s:tw}
Let us now turn attention to relations between unary and binary
two-way automata. It is well known that each \tw\dfa\ can be
converted to an equivalent
\ow\dfa~\cite{RS59,Sh59,Ka05m,GO21,Ch86}\@.
Therefore, by combining these facts with Theorems \ref{t:owbas}
and~\ref{t:decd}, it is easy to see that \iit{a}~each unary
\tw\dfa\ $\cla_1$ can be converted to a binary \ow\dfa~$\cla_2$
(hence, also to binary \tw\dfa) recognizing $\bin\cll(\cla_1)$;
\,\iit{b}~it is decidable whether a given binary \tw\dfa~$\cla_2$
can be converted to a unary $\cla_1$ such that
$\bin\cll(\cla_1)= \cll(\cla_2)$ and, if the answer is positive,
$\cla_1$~can be constructed. We are now going to provide more
direct relations.

\smallbreak
By combining ideas {}from Theorem~\ref{t:swwep} and
Corollary~\ref{c:swwep} with~(\ref{e:owbas}), it is not difficult
to obtain, for each unary \tw\dfa~$\cla_1$, a~\tw\dfa\ recognizing
the binary coded counterpart of~$\cll(\cla_1)$\@: unless the given
binary $w\in \set{0,1}\str$ represents a \quo{small} number, we
can simulate $\cla''_1$ {}from Corollary~\ref{c:swwep}
on~${\vdash}0^N{\dashv}$, where $N= \num w$\@. The automaton
$\cla''_1$ does not use more states than does the
original~$\cla_1$, it is sweeping, in cyclic form with no initial
segment, starting and halting always at the endmarkers. For this
reason, we only have to keep track of states in which $\cla''_1$
visits the endmarkers.
However, $\cla''_1$~may utilize several different entry points
$\fsc q_{i_1},\fsc q_{i_2},\ldots$ to the same loop~$\clc'$\krn,
reached {}from several different states at an endmarker. So, to
avoid a quadratic blow-up in the number of states, we have to
avoid computing the value $(\num w)\mmod\fsc\lambda$ for each of
these entry points separately.

\btheorem{t:twbas}
For each unary \tw\dfa\ $\cla_1$ with $n$~states, there exists a
\tw\dfa~$\cla_2$ recognizing the binary coded counterpart
$\bin\cll(\cla_1)$ with at most $O(n{\cdot}\log n)$ states.

More precisely, if $\cla_1$~uses loops of lengths
$\lambda_1\ldots,\lambda_r$ in the inner graph, $\cla_2$~uses at most
\benumerate
  \item $n{+}2- \sum_{i=1}^{r}\abs{\lambda_i}$ states in a
    preamble deciding membership for \quo{small} binary coded
    values, such that
    \,$\num w\le n- \sum_{i=1}^{r}\abs{\lambda_i}$, and
  \item $\sum_{i=1}^{r}\abs{\lambda_i}{\cdot}(\ell_i{+}2)$ states
    in gadgets simulating the respective loops, where
    $\ell_i\ge 0$ is taken {}from the partial factorization of
    $\lambda_i$ into $\mu_i{\cdot}2^{\ell_i}$ with $\mu_i$~odd.
\eenumerate
This gives
$n{+}2+ \sum_{i=1}^{r}\abs{\lambda_i}{\cdot}(\ell_i{+}1)\le
  n{\cdot}\log n +2n +2$
states in total.
\etheorem
\bproof
Let us fix some notation first:
$n_0= n-\sum_{i=1}^{r}\abs{\lambda_i}=
  n-\sum_{\clc\in\bbc}\abs{\fsc\lambda}$,
where \bbc~is the set of all loops in the inner graph of~$\cla_1$\@.
A~loop $\clc\in \bbc$ is of length
$\fsc\lambda= \fsc\mu{\cdot}2^{\fsc\ell}$\krn, where $\fsc\mu$~is
odd (not necessarily positive) and $\fsc\ell\ge 0$\@. Transition
function for~$\cla_2$ is denoted by~$\delta''$\krn\@.

\smallbreak
$\cla_2$~decides whether $w\in \bin\cll(\cla_1)$ by simulation of
$\cla''_1$ {}from Corollary~\ref{c:swwep} on~${\vdash}0^N{\dashv}$,
where $N= \num w$\@. Since $\cla''_1$ does not have to give a
correct answer if $N= \num w\le n_0$, \,$\cla_2$~decides the
membership in~$\bin\cll(\cla_1)$ for \quo{small} binary coded
values by itself.

This is done by the use of
$\cli= \set{\fsi q_0,\fsi q_1,\ldots,\fsi q_{n_0},\fsi q_{\infty}}$
consisting of $n_0{+}2$ new states, with $\fsi q_0$~as the initial
state. $\cla_2$~computes the numerical value for the prefix which,
so far, has been read {}from the given binary input, in a similar
way as presented by~(\ref{e:owbas}) for one way automata. That is,
for each $e\in \set{0,\ldots,n_0}$ and $b\in \set{0,1}$\@:
\bitemize
  \item $\delta''(\fsi q_0,\vdash)= (\fsi q_0,\rmove)$,
  \item $\delta''(\fsi q_e,b)= (\fsi q_{2e+b},\rmove)$, \,if $2e{+}b\le n_0$,
  \item $\delta''(\fsi q_e,b)= (\fsi q_{\infty},\rmove)$, \,if $2e{+}b> n_0$,
  \item $\delta''(\fsi q_e,\dashv)= \undf$\@.\\
    Depending on whether $0^e\in \cll(\cla_1)$, the state
    $\fsi q_e$ is made accepting or rejecting.
\eitemize

If the computed value exceeds~$n_0$, \,$\cla_2$~switches to the
state~$\fsi q_{\infty}$, in which it traverses the rest of the
input and then it starts to simulate~$\cla''_1$\@. Transitions
for~$\fsi q_{\infty}$ will be given later.

Recall that $\cla''_1$ {}from Corollary~\ref{c:swwep} is sweeping,
in cyclic form with no initial segment, traveling to and fro
across unary inputs by iterated loops.

\bparagraph{Loops, gadgets, and their entry/exit points}
Each loop $\clc'= \set{\fsc q_0,\ldots,\fsc q_{\fsc\lambda-1}}$
of the sweeping unary~$\cla''_1$ is simulated in~$\cla_2$ by a
corresponding \emph{gadget} \clcg\ working on the binary
counterpart of the input. \clcg~contains the original states
$\fsc q_0,\ldots,\fsc q_{\fsc\lambda-1}$ (that is,
$\clc'\subseteq \clcg$), now used as \emph{exit points} {}from the
gadget, then $\fsc r_0,\ldots,\fsc r_{\fsc\lambda-1}$\,---\,new
copies of the original states, used as \emph{entry points} to the
gadget, plus some auxiliary new states, if necessary.

Now, let us describe switching {}from one gadget to another. If
$\cla''_1$ arrives to an endmarker in a state $\fsd q_j$ belonging
to a loop $\cld'\in \bbc$, and then it switches to a state
$\fsc q_i$ belonging to some other loop~$\clc'$, moving the head
one position away {}from this endmarker, then the binary $\cla_2$
will switch {}from $\fsd q_j\in \cldg$ (which is the $j$\mbox{-th}
exit {}from the gadget~\cldg\ simulating~$\cld'$) to the state
$\fsc r_i\in \clcg$ (the $i$\mbox{-th} entry to the gadget~\clcg\
simulating~$\clc'$)\@. This switch moves the head of~$\cla_2$ one
position to the left of~$\dashv$, \emph{regardless} of whether the
original switch {}from $\fsd q_j$ to~$\fsc q_i$ took place
at~$\vdash$ or at~$\dashv$\@. However, if $\cla''_1$~halts
in~$\fsd q_j$, the binary simulator $\cla_2$ halts as well, in the
same state~$\fsd q_j$, with the head parked on~$\dashv$,
preserving acceptance/rejection. We are now ready to introduce
transitions doing this switch: for each loop~$\cld'$ and each
$\fsd q_j\in \cld'$\krn,
\bitemize
  \item $\delta''(\fsd q_j,\dashv)= (\fsc r_i,\lmove)$, \,if\\
    $\delta'(\fsd q_j,\vdash)= (\fsc q_i,\rmove)$ and
      $\cld'$~is a right-to-left loop, or\\
    $\delta'(\fsd q_j,\dashv)= (\fsc q_i,\lmove)$ and
      $\cld'$~is a left-to-right loop,
  \item $\delta''(\fsd q_j,\dashv)= \undf$, \,if\\
    $\delta'(\fsd q_j,\vdash)= \undf$ and $\cld'$~is a
      right-to-left loop, or\\
    $\delta'(\fsd q_j,\dashv)= \undf$ and $\cld'$~is a
      left-to-right loop.\\
    Depending on whether $\fsd q_j\in F_1$, this state is made
       accepting or rejecting in~$\cla_2$\@.
\eitemize

Next, let us describe the internal structure of the gadgets. Recall
that if $\cla''_1$~traverses across~${\vdash}0^N{\dashv}$,
starting in $\fsc q_i\in \clc'$ one position away {}from an
endmarker, it reaches the opposite endmarker in~$\fsc q_k$, where
$k= (i{+}N)\mmod\abs{\fsc\lambda}$\@. The corresponding
gadget~\clcg\ begins in~$\fsc r_i$ and ends in~$\fsc q_k$, which
requires to compute
$k= (i{+}N)\mmod\abs{\fsc\lambda}= (i{+}\num w)\mmod\abs{\fsc\lambda}$,
traversing across~$\vdash w\dashv$ {}from right to left and then back. We
begin by special cases.

\bparagraph{Gadgets for loops with odd lengths}
Consider first a gadget for a loop $\clc'$ of length
$\fsc\lambda= \fsc\mu$, for some odd $\fsc\mu\ge 3$\@.
A~straightforward solution is traversing~$\vdash w\dashv$, to
obtain the value $(\num w)\mmod\fsc\mu$ first and, after that, the
value $i$ is added, modulo~$\fsc\mu$\@. Implemented this way,
$\cla_2$~would use $\Omega(n^2)$ states. However, by the use of
transitions presented by~(\ref{e:owbas}) for one-way automata but,
instead of $j= 0$, starting {}from a value~$j$ satisfying
$(j{\cdot}2^{\abs{w}})\mmod\fsc\mu= i$, we can obtain the value
$(j{\cdot}2^{\abs{w}}+\num w)\mmod\fsc\mu= (i{+}\num w)\mmod\fsc\mu= k$\@.
To obtain such value~$j$, \,$\cla_2$~traverses the entire input to
the left, starting {}from~$\fsc r_e$ with $e= i$\@. At each input
position, the current value $e$ is replaced by a new value
$e'= (e{\cdot}2^{\invcmu})\mmod\fsc\mu$, where%
\footnote{\label{ft:invcmu}It is easy to see that $\invcmu$~does
  exist: by taking the sequence $2^0\!,2^1\!,2^2\!,\ldots,$ we
  must get, sooner or later, two values $g'<g''$ such that
  $2^{g'}\!\mmod\fsc\mu= 2^{g''}\!\mmod\fsc\mu$, and hence
  $2^{g''-g'}\!\mmod\fsc\mu= 1$\@. This gives
  $\invcmu= g''{-}g'{-}1$\@.}
\bequations
  \invcmu &=& \mbox{the smallest nonnegative integer satisfying
    $(2^{\invcmu}{\cdot}2)\mmod\fsc\mu= 1$} \,.
\eequations{e:invcmu}
Thus, after traversing the entire input tape $\vdash w\dashv$ to
the left, $\cla_2$~reaches~$\vdash$ in the state~$\fsc r_e$ with
$e= (\,i{\cdot}(2^{\invcmu})^{\abs{w}}\,)\mmod\fsc\mu$\@. Here
$\cla_2$~switches to the corresponding state~$\fsc q_e$ and
traverses back to the right. At each input position, the current
value $e$~is now replaced in the standard way, in accordance
with~(\ref{e:owbas}), by $e''= (2e{+}b)\mmod\fsc\mu$, where
$b\in \set{0,1}$ denotes the current bit along the input. Thus,
after traversing the entire input back to the right,
$\cla_2$~reaches~$\dashv$ in the state~$\fsc q_e$, with
$e= (\, i{\cdot}(2^{\invcmu})^{\abs{w}}{\cdot}2^{\abs{w}}
    +\num w \,)\mmod\fsc\mu=
  (\, i{\cdot}(2^{\invcmu}{\cdot}2)^{\abs{w}} +\num w \,)\mmod\fsc\mu=
  ( i{\cdot}1 +\num w )\mmod\fsc\mu= k$,
which is the desired exit. Formally, for each loop $\clc'$ of odd
length $\fsc\lambda= \fsc\mu\ge 3$, each
$e\in \set{0,\ldots,\fsc\mu{-}1}$, and each $b\in \set{0,1}$\@:
\bitemize
  \item $\fsc\delta(\fsc r_e,b)= (\fsc r_{e'},\lmove)$,
    \,where $e'= (e{\cdot}2^{\invcmu})\mmod\fsc\mu$,
  \item $\fsc\delta(\fsc r_e,\vdash)= (\fsc q_e,\rmove)$,
  \item $\fsc\delta(\fsc q_e,b)= (\fsc q_{e''},\rmove)$,
    \,where $e''= (2e{+}b)\mmod\fsc\mu$\@.
\eitemize
Here $\fsc\delta$ denotes the function~$\delta''$ with domain
restricted to~\clcg, that is,
$\delta''(p,a)= \fsc\delta(p,a)$, for each $p\in \clcg$ and
$a\in \set{\vdash,0,1,\dashv}$\@. Missing transitions, for
$\fsc q_e$ at~$\dashv$, have already been presented above as
transitions for exits {}from~$\clc'$\krn, switching {}from one gadget
to another.

\bparagraph{Gadgets for loops with lengths equal to powers of two}
Consider now a gadget for a loop $\clc'$ of length
$\fsc\lambda= 2^{\fsc\ell}$\krn, with $\fsc\ell\ge 1$\@.
Also in this case the corresponding
gadget~\clcg\ begins in~$\fsc r_i$ and ends in~$\fsc q_k$, where
$k= (i{+}\num w)\mmod\fsc\lambda$\@. Since here
$\fsc\lambda= 2^{\fsc\ell}$ is a power of two, the value
$(i{+}\num w)\mmod 2^{\fsc\ell}$ depends only on%
\footnote{If $\abs{w}< \fsc\ell$, we handle~$w$ in the same way as
  if it were padded with \quo{sufficiently many} leading zeros\@.}
the last $\fsc\ell$ bits in~$w$\@. More precisely, if
$w= b_{m-1}{\cdots}b_0$ for some bits $b_{m-1},\ldots,b_0$, then,
using the fact that $2^g$ is an integer multiple of $2^{\fsc\ell}$
for each $g\ge \fsc\ell$, we obtain that
$(i{+}\num w)\mmod 2^{\fsc\ell}=
  (i+ \sum_{g=0}^{\fsc\ell-1}b_g{\cdot}2^g)\mmod 2^{\fsc\ell}$\krn\@.
This value is obtained by counting, for
$g=0,\ldots,\fsc\ell{-}1$\@: starting with $e= i$,
\,$\cla_2$~moves along the input to the left and, at each
position, the current value $e$ is replaced by
$e'= (e{+}b_g{\cdot}2^g)\mmod 2^{\fsc\ell}$\krn\@. After passing
through the last $\fsc\ell$ bits in~$w$, the current value $e$
does not change any more; $\cla_2$~just traverses the rest of the
input to the left and then back to~$\dashv$\@. Counting for
$g=0,\ldots,\fsc\ell{-}1$ requires to use new auxiliary states:
besides the exit points
$\fsc q_0,\ldots,\fsc q_{2^{\fsc\ell}-1}\in \clc'$\krn, the gadget
\clcg\ contains~$\fsc r_{g,e}$, for $g=0,\ldots,\fsc\ell$ and
$e= 0,\ldots,2^{\fsc\ell}{-}1$\@. The entry points
$\fsc r_0,\ldots,\fsc r_{2^{\fsc\ell}-1}$ are integrated into
$\fsc r_{0,0},\ldots,\fsc r_{0,2^{\fsc\ell}-1}$, that is,
$\fsc r_0= \fsc r_{0,0}$, $\fsc r_1= \fsc r_{0,1}$, \dots\ Thus,
\clcg~uses $2^{\fsc\ell}{\cdot}(\fsc\ell{+}1)$ new states. This
gives the following transitions, for each loop $\clc'$ of length
$\fsc\lambda= 2^{\fsc\ell}\ge 2$, each
$g\in \set{0,\ldots,\fsc\ell}$, each
$e\in \set{0,\ldots,2^{\fsc\ell}{-}1}$, and each $b\in \set{0,1}$\@:
\bitemize
  \item $\fsc\delta(\fsc r_{g,e},b)= (\fsc r_{g+1,e'},\lmove)$,
    \,where $e'= (e{+}b{\cdot}2^g)\mmod 2^{\fsc\ell}$,
    \,if $g< \fsc\ell$,
  \item $\fsc\delta(\fsc r_{\fsc\ell,e},b)= (\fsc r_{\fsc\ell,e},\lmove)$,
  \item $\fsc\delta(\fsc r_{g,e},\vdash)= (\fsc q_e,\rmove)$,
  \item $\fsc\delta(\fsc q_e,b)= (\fsc q_e,\rmove)$\@.
\eitemize
Again, $\fsc\delta$ denotes the function~$\delta''$ with domain
restricted to the states in~\clcg\ and missing transitions for
$\fsc q_e$ at~$\dashv$ have already been presented, as transitions
for exit points.

\bparagraph{Gadgets for loops of length one}
This time we need a gadget for a loop $\clc'$ of length
$\fsc\lambda= 1$\@. This length
can be used as the smallest odd length, that is, if
$\fsc\lambda= \fsc\mu= 1$, as well as the smallest power of two,
that is, if $\fsc\lambda= 2^{\fsc\ell}$ with $\fsc\ell= 0$\@. The
corresponding gadget \clcg\ consists only of two states, namely,
$\fsc q_0$ and~$\fsc r_0$, the only exit and entry points. Here we
actually do not have to compute anything, since
$(i{+}\num w)\mmod 1= 0$ for each $i$ and~$w$\@. Nevertheless, to
keep the trajectory of input head movement uniform,
we do traverse the entire input to the left and then back: for
each loop $\clc'$ of length $\fsc\lambda= 1$ and each
$b\in \set{0,1}$\@:
\bitemize
  \item $\fsc\delta(\fsc r_0,b)= (\fsc r_0,\lmove)$,
    \ $\fsc\delta(\fsc r_0,\vdash)= (\fsc q_0,\rmove)$,
    \ $\fsc\delta(\fsc q_0,b)= (\fsc q_0,\rmove)$\@.
\eitemize

\bparagraph{Gadgets for loops moving to the right}
Next, consider a gadget for a loop $\clc'$ of length
$\fsc\lambda= \fsc\mu{\cdot}2^{\fsc\ell}$\krn, with odd
$\fsc\mu\ge 1$ and $\fsc\ell\ge 0$\@. The loop
$\clc'= \set{\fsc q_0,\fsc q_1,\ldots,\fsc q_{\fsc\lambda-1}}$ can
be decomposed to a Cartesian product of two simpler loops, namely,
to $\clc'= \cld'{\times}\cle'$\krn, where
$\cld'= \set{\fsd q_0,\ldots,\fsd q_{\fsc\mu-1}}$ and
$\cle'= \set{\fse q_0,\ldots,\fse q_{2^{\fsc\ell} -1}}$\@. These
two loops are of lengths $\fsc\mu$ and~$2^{\fsc\ell}$\krn,
respectively. So far, we have not introduced any new states, just
each state $\fsc q_i\in \clc'$ can also be viewed as
$\fsc q_i= \angl{\fsd q_{i{\bmod}\fsc\mu},\fse q_{i{\bmod}2^{\fsc\ell}}}\in
  \cld'{\times}\cle'$\krn\@.
By the Chinese Remainder Theorem, such mapping is unambiguous.

Now, by the use of constructions given above for simpler loops, we
first construct two preliminary drafts, the corresponding gadgets
$\cldg$ and~$\cleg$, with the respective transition functions
$\fsd\delta$ and~$\fse\delta$\@. This gives us also some new
states, among others, the entry points
$\fsd r_0,\ldots,\fsd r_{\fsc\mu-1}\in \cldg$ and
$\fse r_0,\ldots,\fse r_{2^{\fsc\ell}-1}\in \cleg$\@. Having this
done, we can utilize $\cldg{\times}\cleg$ to obtain a gadget for
the loop~$\clc'$\@. We have already the exit points:
$\fsc q_i= \angl{\fsd q_{i{\bmod}\fsc\mu},\fse q_{i{\bmod}2^{\fsc\ell}}}$
belongs to $\clc'= \cld'{\times}\cle'\subseteq \cldg{\times}\cleg$\@.
Now, by definition, let
$\fsc r_i= \angl{\fsd r_{i{\bmod}\fsc\mu},\fse r_{i{\bmod}2^{\fsc\ell}}}$,
for each $i= 0,\ldots,\fsc\lambda{-}1$, which establishes the
entry points in $\cldg{\times}\cleg$\@.

The important fact is that all simpler gadgets move the input head
in the same way and visit the endmarkers at the same moments of
time. This allows to simulate the computations of $\cldg$
and~$\cleg$ in parallel, sharing the same input head.
Transitions for this task are straightforward: for each
$\fsd p\in \cldg$, $\fse p\in \cleg$, and $a\in \set{\vdash,0,1}$,
\bitemize
  \item $\fsc\delta(\angl{\fsd p,\fse p},a)= (\angl{\fsd p',\fse p'},d)$,
    \,if $\fsd\delta(\fsd p,a)= (\fsd p',d)$ and
    $\fse\delta(\fse p,a)= (\fse p',d)$\@.
\eitemize
Also in this case the missing transitions, with $a= {\dashv}$,
have already been presented above, as transitions for exit points
{}from~$\clcg$.
It should be pointed out that some states in $\cldg{\times}\cleg$
are not reachable {}from entry points, since $\cla_2$ moves to the
left in states belonging to
$(\cldg\setminus\cld'){\times}(\cleg\setminus\cle')$, and then
back in states belonging to $\cld'{\times}\cle'$\krn\@. The
reachable part of $\cldg{\times}\cleg$ thus consists of at most
$\fsc\mu{\times}2^{\fsc\ell}{\cdot}(\fsc\ell{+}1) +
    \fsc\mu{\times}2^{\fsc\ell}=
  \fsc\mu{\cdot}2^{\fsc\ell}{\cdot}(\fsc\ell{+}2)=
  \fsc\lambda{\cdot}(\fsc\ell{+}2)$
states.

\bparagraph{Gadgets for loops not moving to the right}
Since the sweeping $\cla''_1$ {}from Corollary~\ref{c:swwep} does
not use loops of length $\fsc\lambda= 0$, this case leaves us with
a loop $\clc'$ of length $\fsc\lambda< 0$\@. The corresponding
gadget \clcg\ is implemented in the same way as gadgets moving to
the right, even with the same input head movement, but all
values are computed modulo~${-}(\fsc\lambda)$\@.

\bparagraph{Activation of the first gadget}
Recall that, in the initial phase, $\cla_2$~enters the
state~$\fsi q_{\infty}$ if it finds that $N= \num w$ is
sufficiently large for simulation of~$\cla''_1$
on~${\vdash}0^N{\dashv}$\@. Recall also that the initial state
of~$\cla''_1$ is some~$\fsc q_h$, belonging to a left-to-right
loop~$\clc'$\krn, and that the first executed transition is
$\delta'(\fsc q_h,\vdash)= (\fsc q_h,\rmove)$, after which
$\cla''_1$~starts iteration of~$\clc'$\krn\@. This leads to the
following transitions for~$\fsi q_{\infty}$\@:
\bitemize
  \item $\delta''(\fsi q_{\infty},b)= (\fsi q_{\infty},\rmove)$,
    \,for each $b\in \set{0,1}$,
  \item $\delta''(\fsi q_{\infty},\dashv)= (\fsc r_h,\lmove)$,
    \,where $\fsc r_h$~is the entry point in the gadget~\clcg\
    corresponding, in~$\cla''_1$, to the initial state~$\fsc q_h$
    in the loop~$\clc'$\krn\@.
\eitemize
After that, $\cla_2$~proceeds by simulation of the first traversal
of~$\cla''_1$ across~${\vdash}0^N{\dashv}$\@.

\smallbreak
The correctness of $\cla_2$ follows {}from the correctness of
gadgets, by induction on the number of visits at the endmarkers
of~${\vdash}0^N{\dashv}$\@.
The upper bound on the number of states is straightforward: the
preamble deciding membership for \quo{small} binary inputs uses
$n_0{+}2= n{+}2 -\sum_{\clc\in\bbc}\abs{\fsc\lambda}$ states; the
gadgets simulating the respective loops
$\sum_{\clc\in\bbc}\abs{\fsc\lambda}{\cdot}(\fsc\ell{+}2)$ states.
Clearly, $\fsc\ell\le \log n$ for each
$\clc\in \bbc$, or else, for some~$\clc$, we get
$\abs{\fsc\lambda}= \abs{\fsc\mu}{\cdot}2^{\fsc\ell}> 2^{\log n}= n$,
a contradiction. This gives
$n{+}2 +\sum_{\clc\in\bbc}\abs{\fsc\lambda}{\cdot}(\fsc\ell{+}1)\le
  n{+}2 + n{\cdot}(\log n{+}1)$
states in total.
\eproof

It is not known whether the upper bound $O(n{\cdot}\log n)$ given
by Theorem~\ref{t:twbas} cannot be improved. However, if the
original unary \tw\dfa\ $\cla_1$ does not have loops of even
lengths in the inner graph (which implies that the period of
$\cll(\cla_1)$ is odd), the above upper bound drops down to
$2n{+}2$ states, since then $\fsc\ell =0$ for each loop~$\clc$\@:

\bcorollary{c:twbas}
For each unary \tw\dfa\ $\cla_1$ with $n$~states, using only loops
of odd lengths in the inner graph, there exists a binary
\tw\dfa~$\cla_2$ recognizing $\bin\cll(\cla_1)$ with at most
$2n{+}2$ states.
\ecorollary

To see that the linear upper bound {}from Corollary~\ref{c:twbas}
cannot be improved in the case of unary \tw\dfas\ using only loops
of odd lengths in the inner graph, we are now going to show a
conversion in the opposite way, {}from binary \tw\dfas\ to unary
\tw\dfas\@:

\btheorem{t:unry}
Let $\cla_2$ be a binary \tw\dfa\ with $n$~states. If
$\cll(\cla_2)$ is a binary coded version of a unary
language~\cll\ with no tail and an odd period, then there exists
a unary \tw\dfa~$\cla_1$ recognizing~\cll\ with at most $n$~states.
\etheorem
\bproof
First, given a binary \tw\dfa~$\cla_2$, it is decidable whether
$\cll(\cla_2)= \bin\cll$ for some unary language~\cll\ with no
tail and an odd period~$\lambda$ and, if the answer is positive,
the value $\lambda$~can be computed: first, we can convert
$\cla_2$ to an equivalent \ow\dfa\ $\cla'_2$~\cite{GO21} and
then, by Theorem~\ref{t:decd} (see
also~\cite[Cor.~14, Alg.~1]{Ge25}), we can decide whether there
exists a corresponding unary \ow\dfa~$\cla'_1$ such that
$\cll(\cla_2)= \cll(\cla'_2)= \bin\cll(\cla'_1)$\@. Moreover, if
the answer is positive, the algorithm constructs~$\cla'_1$ that is
minimal, which gives us an initial segment of length $\sigma\ge 0$
and a loop of length $\lambda\ge 1$\@. Finally, we verify whether
$\sigma= 0$ and $\lambda$ is odd.%
\footnote{If the given $\cla_2$ does not pass some of these tests,
  we do not construct~$\cla_1$\@.}

In what follows, we shall also assume that $\lambda\ne 1$\@: if
$\lambda= 1$ and $\sigma= 0$, then either $\cll(\cla'_1)= \emset$
or $\cll(\cla'_1)= 0\str$ and $\cll= \cll(\cla'_1)$ can be recognized by
the use of a single state.

\smallbreak
Now, after obtaining~$\lambda$ (handled as a~fixed constant {}from
now on), let $i_{\lambda}$~be the smallest nonnegative integer
satisfying $(2^{i_{\lambda}}{\cdot}2)\mmod\lambda= 1$
(see also~(\ref{e:invcmu}) and Footnote~\ref{ft:invcmu})\@.

For each $N\ge 0$, consider now the binary input
$w\xss{N}= (0^{i_{\lambda}}1)^{N+2\lambda}$\@. Since
$\cll(\cla_2)= \bin\cll$, we have that $w\xss{N}\in \cll(\cla_2)$
if and only if $0^{\num w\xss{N}}\in \cll$\@. But \cll~has the
period~$\lambda$ with no tail, and hence
$0^{\num w\xss{N}}\in \cll$ if and only if
$0^{(\num w\xss{N}){\bmod}\lambda}\in \cll$\@. It is easy to see
that the binary string $w\xss{N}$ represents the number
$\num w\xss{N}= \sum_{k=0}^{N+2\lambda-1} 2^{(i_{\lambda}+1)\cdot k}$\krn\@.
Taken this value modulo~$\lambda$, we get
$(\num w\xss{N})\mmod\lambda=
  (\sum_{k=0}^{N+2\lambda-1}(2^{i_{\lambda}}{\cdot}2)^k)\mmod\lambda=
  (\sum_{k=0}^{N+2\lambda-1} 1)\mmod\lambda=
  (N{+}2\lambda)\mmod\lambda= N\mmod\lambda$\@.
Thus, $0^{(\num w\xss{N}){\bmod}\lambda}\in \cll$ if and only
if $0^{N{\bmod}\lambda}\in \cll$, which in turn holds if and only
if $0^N\in \cll$\@. To sum it up, $0^N\in \cll$ if and only if
$w\xss{N}\in \cll(\cla_2)$\@.

For this reasons, the unary \tw\dfa~$\cla_1$, using the same
states as does~$\cla_2$, can decide whether $0^N\in \cll$ by
simulating~$\cla_2$ on~$w\xss{N}$\@. That is, the unary input
${\vdash}0^N{\dashv}$ is interpreted as
${\vdash}w\xss{N}{\dashv}= {\vdash}(0^{i_{\lambda}}1)^{\lambda}{\cdot}
  (0^{i_{\lambda}}1)^{N}{\cdot} (0^{i_{\lambda}}1)^{\lambda}{\dashv}$\@.
More precisely, if the head of $\cla_1$ is reading a
symbol~\quo{$0$} at some position along~${\vdash}0^N{\dashv}$,
\,$\cla_1$~pretends that $\cla_2$~is reading~\quo{$1$} in the
middle of~$\ldots 0^{i_{\lambda}}10^{i_{\lambda}}\ldots$ at the
corresponding position along~${\vdash}w\xss{N}{\dashv}$\@.
A~segment of the original computation path of~$\cla_2$ that starts
at this symbol~\quo{$1$} and, after
leaving~$0^{i_{\lambda}}10^{i_{\lambda}}$, it reaches the nearest
symbol~\quo{$1$} to the left/right, is simulated by a single-step
transition along~${\vdash}0^N{\dashv}$\@. For technical reasons, to
avoid problems with acceptance/rejection when the simulation ends,
the endmarkers are handled as follows:
if the head of $\cla_1$ is reading~$\vdash$, \,$\cla_1$~pretends
that $\cla_2$~is reading the rightmost~\quo{$1$} in
${\vdash}(0^{i_{\lambda}}1)^{\lambda}0^{i_{\lambda}}$, similarly,
if the head is reading~$\dashv$, \,$\cla_1$~pretends that
$\cla_2$~is reading the leftmost~\quo{$1$} in
$(0^{i_{\lambda}}1)^{\lambda}{\dashv}$\@. This leads to the
following transition function~$\delta'$ and to the following
initial state~$q'\xsr{I}$\@:
\bitemize
  \item $\delta'(q,\vdash)= (q',\rmove)$,
    \,if the computation of~$\cla_2$, starting {}from~$q$ with the
    head on the rightmost symbol~\quo{$1$} of the string
    $v\xss{\vdash}= {\vdash}(0^{i_{\lambda}}1)^{\lambda}0^{i_{\lambda}}$,
    leaves~$v\xss{\vdash}$ in the state~$q'$\krn\@.
  \item $\delta'(q,0)= (q',d)$,
    \,if the computation of~$\cla_2$, starting {}from~$q$ with the
    head on the symbol~\quo{$1$} of the string
    $v\xss{0}= 0^{i_{\lambda}}10^{i_{\lambda}}$, leaves~$v\xss{0}$
    in the state~$q'$\krn\@. Depending on whether $\cla_2$
    leaves~$v\xss{0}$ to the left or to the right, $d= \lmove$ or
    $d= \rmove$\@.
  \item $\delta'(q,\dashv)= (q',\lmove)$,
    \,if the computation of~$\cla_2$, starting {}from~$q$ with the
    head on the leftmost symbol~\quo{$1$} of the string
    $v\xss{\dashv}= (0^{i_{\lambda}}1)^{\lambda}{\dashv}$,
    leaves~$v\xss{\dashv}$ in the state~$q'$\krn\@.
\smallbreak
  \item $\delta'(q,a)= \undf$,
    \,if the computation of~$\cla_2$, starting on the string
    $v_a\in \set{v\xss{\vdash},v\xss{0},v\xss{\dashv}}$ {}from~$q$
    with the head on the corresponding symbol~\quo{$1$} (the
    rightmost~\quo{$1$} for $v_a= v\xss{\vdash}$, the
    only~\quo{$1$} for $v_a= v\xss{0}$, and the leftmost~\quo{$1$}
    for $v_a= v\xss{\dashv}$), does not leave~$v_a$, but halts.
    Depending on whether $\cla_2$ halts in $q'\in F$ or
    $q'\notin F$, the state $q$ is made accepting or rejecting.
  \item $\delta'(q,a)= \undf$,
    \,if the computation of~$\cla_2$, starting on the string
    $v_a\in \set{v\xss{\vdash},v\xss{0},v\xss{\dashv}}$ {}from~$q$
    with the head on the corresponding symbol~\quo{$1$} (specified
    as above), does not leave~$v_a$, nor does it halt in
    $n{\cdot}\abs{v_a}$ steps\@.
    The state $q$ is made rejecting.
\smallbreak
  \item $q'\xsr{I}= q$,
    \,where $q$~is the first state in which, starting {}from the
    initial state at the left endmarker, $\cla_2$~gets to the
    rightmost symbol~\quo{$1$} of the string~$v\xss{\vdash}$\@.
\eitemize

In the first step, by $\delta'(q'\xsr{I},\vdash)= (q',\lmove)$,
\,$\cla_1$~gets to the state~$q'$ in which, starting {}from the
initial state at the left endmarker, $\cla_2$~leaves
$v\xss{\vdash}= {\vdash}(0^{i_{\lambda}}1)^{\lambda}0^{i_{\lambda}}$
to the right for the first time. It should be pointed out that
$q'\xsr{I}= q$ and~$q'$ do exist: if $\cla_2$~never
leaves~$v\xss{\vdash}$ to the right, the language
$\cll(\cla_2)\cap(0^{i_{\lambda}}1)\str$ is either finite or
cofinite, which implies that \cll~is either finite or cofinite,
since $0^N\in \cll$ if and only if $w\xss{N}\in \cll(\cla_2)$\@.
This gives the period $\lambda= 1$ for~\cll, the case we have
eliminated already.

\smallbreak
Note also that, even though the states of~$\cla_1$ are made
accepting or rejecting for several different reasons, the
acceptance/rejection is set unambiguously:

Consider the case of $\delta'(q,\vdash)= \undf$, with the
state~$q$ made accepting. If the state $q$ is required to be
accepting because, starting {}from~$q$ at the rightmost~\quo{$1$} of
the string
$v\xss{\vdash}= {\vdash}(0^{i_{\lambda}}1)^{\lambda}0^{i_{\lambda}}$,
\,$\cla_2$~halts in some $q'\in F$ not moving farther than
$i_{\lambda}$~positions away (which claims
$\delta'(q,\vdash)= \undf$), then $\cla_2$~halts in $q'\in F$ not
moving farther than $i_{\lambda}$~positions away even if the
computation starts {}from~$q$ at the only~\quo{$1$} of
$v\xss{0}= 0^{i_{\lambda}}10^{i_{\lambda}}$ or at
the leftmost~\quo{$1$} of
$v\xss{\dashv}= (0^{i_{\lambda}}1)^{\lambda}{\dashv}$\@. This
claims $\delta'(q,0)= \undf$, \,$\delta'(q,\dashv)= \undf$, and
hence, two times, that $q$~must be accepting.

On the other hand, if $q$~is required to be accepting because,
starting {}from~$q$ at the rightmost~\quo{$1$} of the string
$v\xss{\vdash}= {\vdash}(0^{i_{\lambda}}1)^{\lambda}0^{i_{\lambda}}$,
\,$\cla_2$~halts in an accepting state after moving farther than
$i_{\lambda}$~positions to the left, but without
leaving~$v\xss{\vdash}$ (which claims $\delta'(q,\vdash)= \undf$),
then the computation path must pass through a state~$q'$ in which
$\cla_2$~gets $i_{\lambda}$~positions to the left for the first
time. This claims $\delta'(q,0)= (q',\lmove)$,
\,$\delta'(q,\dashv)= (q',\lmove)$, with no demands about
acceptance/rejection on~$q$\@.

A~similar reasoning holds for rejection\,---\,by halting in
a rejecting state or by en\-ter\-ing an infinite cycle, as well as
for the cases of $\delta'(q,0)= \undf$ or
$\delta'(q,\dashv)= \undf$\@.
\eproof

The construction presented in Theorem~\ref{t:unry} above can be
used for any $\cla_2$ recognizing a binary coded version of any
unary language~\cll\ with an odd period, even if the tail is of
length $\sigma> 0$\@. However, since the constructed
$\cla_1$ works correctly under the condition that
$0^{N{\bmod}\lambda}\in \cll$ if and only if $0^N\in \cll$, which
here does not necessarily hold for $N< \sigma$, we can only grant
that $\cla_1$~agrees with~\cll\ on inputs of length
$N\ge \sigma$\@. This can be fixed by using a \tw\dfa\
$\cla'_1$ with additional $\sigma{+}1$ states, which decides the
membership in~\cll\ for \quo{short} inputs by itself and
simulates~$\cla_1$ only on inputs of length $N\ge \sigma$\@.
Summing up, if $\cla_2$~uses $n$~states
and $\cll(\cla_2)$~is a binary coded version of a unary
language~\cll\ with a tail~$\sigma$ and an odd period~$\lambda$,
then there exists a unary \tw\dfa\ $\cla'_1$ recognizing~\cll\ with
at most $n{+}\sigma{+}1$ states.

Next, as a direct consequence of Theorem~\ref{t:unry} for unary
languages with no tail and an odd period, the linear upper bound
{}from Corollary~\ref{c:twbas} cannot be asymptotically improved:

\bcorollary{c:oddlow}
Let $\cla_1$ be a minimal unary \tw\dfa\ using $n$~states, all of
them in loops of odd lengths in the inner graph. Then any binary
\tw\dfa\ recognizing $\bin\cll(\cla_1)$ must use at least
$n$~states.
\ecorollary
\bproof
Let $\lambda_1\ldots,\lambda_r$ be the lengths of all loops in the
inner graph of~$\cla_1$, and let
$\lambda= \lcm\set{\abs{\lambda_1},\ldots,\abs{\lambda_r}}$ be the
least common multiple of these values. All loops are of odd
lengths, and hence $\lambda$~must also be odd.

It is well known that $\lambda$~must be a period for
$\cll(\cla_1)$~\cite{KO11,KO12}\@. The argument for this is quite
straightforward: by assumptions of the theorem, $\cla_1$~can
travel {}from one endmarker to another only by iterating a loop. But
a~loop of length $\lambda_i\ne 0$ can be iterated
$\lambda/\abs{\lambda_i}$ more times, which travels exactly
$\lambda$ additional positions, in the same direction. Therefore,
for each $N\ge 0$, \,$\cla_1$ visits the endmarkers on
${\vdash}0^N{\dashv}$ and on~${\vdash}0^{N+\lambda}{\dashv}$ by
the same sequence of states, and hence $0^N\in \cll(\cla_1)$ if
and only if $0^{N+\lambda}\in \cll(\cla_1)$\@. Thus,
$\cll(\cla_1)$ is a unary language with no tail and an odd period.

Suppose now, for contradiction, that $\bin\cll(\cla_1)$ can be
recognized by a binary \tw\dfa\ with $n'< n$ states. But then, by
Theorem~\ref{t:unry}, there exists a unary \tw\dfa\
recognizing~$\cll(\cla_1)$ with at most $n'$~states, which
contradicts the fact that $\cla_1$~is minimal.
\eproof

The next theorem provides witness automata satisfying the
assumptions of Corollary~\ref{c:oddlow} above. This gives, for
each $n\ge 7$, some minimal unary \tw\dfas\ for which conversion
to binary counterparts cannot save a single state.

\btheorem{t:oddgap}
For each $n\ge 7$, there exists a unary language~$\cll_n$ for
which a minimal \tw\dfa\ uses exactly $n$~states such that any
minimal binary \tw\dfa\ recognizing $\bin\cll_n$ must use at least
$n$~states, but less than $n +\log n -1$ states.
\etheorem
\bproof
First, by Lemma~\ref{l:primes}, each $n\ge 7$, except for $n= 9$,
can be expressed as a sum of less than $\log n$ odd primes, all
different. (The case of $n= 9$ will be discussed later\@.) So let
us begin with partitioning the given number~$n$ to a~sum
$n = \sum_{i=1}^r p_i$, where $p_1,\ldots,p_r$ are pairwise
distinct odd primes, with $r< \log n$, and let
$p = \prod_{i=1}^r p_i$\@. Consider now
\bdisplay
  \cll_n &=& \set{0^N\st N\mmod p= 0} \,.
\edisplay

It is quite obvious that the minimal \ow\dfa\ recognizing~$\cll_n$
uses exactly $p = \prod_{i=1}^r p_i$ states, all of them in
a single loop, counting modulo~$p$, with no initial segment. But
then, by Theorem~B in~\cite{KO12} (see also~\cite{KO11}), any
\tw\dfa\ recognizing~$\cll_n$ must use at least $\sum_{i=1}^r p_i$
states.

Next, for \tw\dfas, it is easy to see that $n= \sum_{i=1}^r p_i$
states are also sufficient: to decide whether $N\mmod p= 0$, our
\tw\dfa\ $\cla_1$~verifies, one after another, whether $N$~is
divisible by~$p_i$, for $i= 1,\ldots,r$, alternating between
left-to-right and right-to-left traversals along
${\vdash}0^N{\dashv}$\@. Therefore, any minimal \tw\dfa\
for~$\cll_n$ uses exactly $n$~states.

Clearly, the inner graph of~$\cla_1$ consists of~$n$ states,
grouped into $r$~loops of lengths $\lambda_1,\ldots,\lambda_r$,
where $\lambda_i= {+}p_i$ or $\lambda_i= {-}p_i$, depending on
whether $i$~is odd or even. But then, by Corollary~\ref{c:oddlow},
any binary \tw\dfa\ recognizing $\bin\cll(\cla_1)= \bin\cll_n$ must
use at least $n$~states.

On the other hand, a \tw\dfa\ for~$\bin\cll_n$ can be constructed
quite easily: it is enough to verify, for $i= 1,\ldots,r$, one
after another, whether the given binary number is divisible
by~$p_i$\@. This can be done in a similar way as presented
by~(\ref{e:owbas}) for one way automata, which only requires
$r{-}1$ additional states, to return the head back the left
endmarker in between two left-to-right traversals along the binary
input. The total number of states is thus bounded by
$\sum_{i=1}^r p_i +(r{-}1)< n+\log n -1$\@.

The above reasoning works also for partitioning of~$n$ to a sum of
prime powers, i.e., for $n = \sum_{i=1}^r p_i^{\alpha_i}$\krn,
where $p_1,\ldots,p_r$ are pairwise distinct odd primes and
$\alpha_1,\ldots,\alpha_r$ are positive integers. This gives an
argument for $n= 9$, since $n= 9= 3^2$\krn\@.
\eproof

\bsection{Concluding Remarks}{s:conc}
We have shown that each unary \tw\dfa\ with $n$~states can be
transformed to a \tw\dfa~recognizing the binary coded counterpart of
the original language with at most $O(n{\cdot}\log n)$ states. It
is not known whether this upper bound cannot be improved. However,
we have presented unary witness \tw\dfas\ for which at least $n$
states are necessary.

The linear (or close to linear) relations between unary and binary
\tw\dfas, presented by Theorems \ref{t:unry} and~\ref{t:oddgap}
as well as by Corollaries \ref{c:twbas} and~\ref{c:oddlow}, were
established for unary regular languages with odd periods. For
example, the construction given in Theorem~\ref{t:unry} does not
work for a language with a period that is even, since then there
is no positive integer~$i_{\lambda}$ satisfying
$(2^{i_{\lambda}}{\cdot}2)\mmod\lambda= 1$\@. 

We do not know whether the number of primes forming the sum in
Lemma~\ref{l:primes} can be reduced. This could improve the upper
bound given by Theorem~\ref{t:oddgap}\@.

We are also convinced that several statements presented in
Section~\ref{s:tw} hold for nondeterministic two-way automata as
well.

%
\end{document}